# Смертность от болезней системы кровообращения и болезней органов дыхания, ассоциированная с гриппом в Российской Федерации во время сезонов гриппа с 2013-2014 до 2018-2019.

Эдвард Гольдштейн[1,*]

1. Гарвардская школа общественного здравоохранения, Бостон, США
*. Электронная почта: egoldste@hsph.harvard.edu

**Аннотация**

*Актуальность*: Информация о смертности, ассоциированной с гриппом в России ограничена и в основном базируется на диагностированных случаях смерти от гриппа, которые представляют только малую долю всех смертей, ассоциированных с гриппом. *Цель*: Оценить смертность от болезней системы кровообращения и болезней органов дыхания, ассоциированную с гриппом в Российской Федерации во время сезонов гриппа 2013/14 до 2018/19. *Материалы и Методы*: Используя ранее разработанную регрессионную модель, мы выразили месячные уровни смертности в Российской Федерации от болезней органов дыхания, а также болезней системы кровообращения между 07/2013 и 07/2019 (данные Росстата) через месячные индикаторы циркуляции гриппа A/H3N2, A/H1N1 и B (полученные из данных института гриппа им. Смородинцева), базовые (с годовой периодичностью) уровни месячной смертности не связанной с гриппом и временной тренд в смертности. *Результаты и обсуждение*: В сезоны 2013/14 до 2018/19, в среднем 17636 (95% ДИ (9482,25790)) годовых смертей от болезней системы кровообращения и 4179 (3250,5109) смертей от болезней органов дыхания были ассоциированы с гриппом. Наибольшая смертность как от болезней системы кровообращения (32298 (18071,46525)), так и от болезней органов дыхания (6689 (5019,8359)), ассоциированная с гриппом была оценена в сезон 2014/15. Среди смертей, ассоциированных с гриппом, грипп A/H3N2 был причиной 51.8% смертей от болезней системы кровообращения и 37.2% смертей от болезней органов дыхания; грипп A/H1N1 был причиной 23.4% смертей от болезней системы кровообращения и 29.5% смертей от болезней органов дыхания; грипп B был причиной 24.9% смертей от болезней системы кровообращения и 33.3% смертей от болезней органов дыхания, подавляющее большинство которых было связанно с гриппом B/Ямагата. По сравнению с сезонами 2013/14 до 2015/16, в сезоны 2016/7 до 2018/19 (когда уровень вакцинации против гриппа значительно возрос), смертность от гриппа упала на 16.1%, или 3809 ежегодных смертей от болезней системы кровообращения и болезней органов дыхания. **Выводы**: Циркуляция гриппа связанна с существенной смертностью, особенно от болезней системы кровообращения в Российской Федерации; эта смертность несколько понизилась после увеличения уровня вакцинации против гриппа. Эти результаты поддерживают потенциальную важность дополнительного увеличения уровня вакцинации против гриппа, особенно среди людей с сердечно-сосудистыми заболеваниями и пожилых людей, использование четырехвалентной вакцины против гриппа, а также применение антивирусных препаратов в определенных группах населения во время активной циркуляции гриппа.

**Введение**

Основные подтипы гриппа (A/H3N2, A/H1N1 и B) ежегодно циркулируют в Российской Федерации [1,2]. Известно, что подобные эпидемии приводят к значительному уровню смертности в странах северного полушария [3-9], включая смертность от болезней системы кровообращения [3,4]; известно также, что инфекция гриппа связана с рядом сердечно-сосудистых осложнений [10,11]. В то же время, информация о вкладе гриппа в смертность в Российской Федерации (как и в общую смертность, так и в смертность от различных классов болезней, в частности болезней системы кровообращения) ограничена. Ежегодные оценки количества смертей с диагностированным гриппом (например [2,12-15]) предположительно отражают только малую долю (несколько процентов) всех смертей, ассоциированных с гриппом (т.е. смертей, для которых инфекция гриппа является триггером, а не только смертей, при которых грипп диагностирован), особенно среди старших возрастных групп. Например в [11] показано, что лабораторно диагностированная инфекция гриппа повышает риск острого инфаркта миокарды приблизительно в 6 раз; при этом, для избыточных инфарктов и связанной сними смертностью при эпидемиях гриппа, инфекция гриппа диагностируется крайне редко. Отметим также, что уровень смертности от болезней системы кровообращения в России высок, и что при эпидемиях гриппа, скачки в уровне смертности от болезней системы кровообращения по абсолютной величине заметно превышают скачки в уровне смертности от болезней органов дыхания (Рис. 1,3,4). В дополнении к этому, случаи смерти с диагностированным гриппом могут неверно отражать относительный вклад разных подтипов гриппа (A/H3N2, A/H1N1 и B) в смертность, ассоциированную с гриппом. При эпидемиях гриппа A/H1N1, распределение по возрасту у смертей с диагностированным гриппом значительно моложе, чем при эпидемиях гриппа A/H3N2 и B [13]. Соответственно, смерти, ассоциированные с гриппом A/H1N1 чаще диагностируемы, чем смерти, ассоциированные с гриппом A/H3N2 и B. Действительно, грипп A/H1N1 был самой распространенной причиной диагностированных смертей с гриппом даже в сезон 2014/15, когда циркуляция гриппа A/H1N1 была мала [14,15]. Все это говорит о том, что данные о смертях с диагностированным гриппом не только сильно недооценивают общий уровень смертности, ассоциированной с гриппом, но и потенциально (а) искажают соотношение уровня смертности, ассоциированной с гриппом в различные сезоны, переоценивая относительный вклад гриппа A/H1N1 в общую смертность от гриппа; (б) недооценивают долю смертей от болезней системы кровообращения среди всех смертей, ассоциированных с гриппом; (в) недооценивают долю пожилых людей среди всех смертей, ассоциированных с гриппом, и т.д. Соответственно, необходима иная методология чем изучение смертей с диагностированным гриппа [2,12-15] для того, чтобы лучше оценить бремя смертности от разных болезней, ассоциированной с основными подтипами гриппа (A/H3N2, A/H1N1, B) в разные годы, и лучше понять эффект недавнего увеличения уровня вакцинации против гриппа в РФ на смертность, ассоциированную с гриппом.

В наших ранних работах [3,4,16] мы ввели статистический метод для оценки уровней смертности и госпитализаций, ассоциированных с гриппом и респираторно-синцитиальным вирусом (РС-вирусом), разработанный с целью устранения ряда ограничений в предыдущей методологии оценки уровня смертности и госпитализаций от респираторных вирусов. Один из важных аспектов этой методологии является использование "индексов" циркуляции гриппа A/H3N2, A/H1N1 и B, а также РС-вируса которые пропорциональны уровням инфекций с соответствующими вирусами в населении. Для гриппа, эти (недельные, или месячные) индексы строятся из данных о медицинских консультациях с симптомами гриппа/ОРВИ, а также данных о тестировании респираторных образцов у людей с симптомами гриппа/ОРВИ на разные подтипы гриппа [3,4,16]. Мы использовали индексы циркуляции основных подтипов гриппа (A/H3N2, A/H1N1 и B) для того, чтобы оценить смертность, ассоциированную с гриппом в разных возрастных группах для различных причин смерти в США [3,4]. В последствии, этот метод был применен для оценки уровня смертности, ассоциированной с гриппом в ряде других стран [6-9]. В этой статье, мы определили и вычислили аналогичные индексы циркуляции гриппа используя эпидемиологические данные института гриппа им. Смородинцева [17,18] о медицинских консультациях с симптомами гриппа/ОРВИ, о тестировании респираторных образцов у людей с симптомами гриппа/ОРВИ на разные подтипы гриппа, и о антигенной характеризации полученных образцов гриппа. Эти индексы циркуляции гриппа были использованы в этой статье в сочетании с месячными данными Росстата о смертности [19] в рамках статистической модели, разработанной в [3,4] для того, чтобы оценить смертность, ассоциированную с гриппом в сезоны 2013/14 до 2018/19 в РФ.

*Цель исследования:* Оценка смертности от болезней системы кровообращения и болезней органов дыхания, ассоциированной с гриппом в Российской Федерации во время сезонов гриппа с 2013/14 до 2018/19, включая относительный вклад гриппа A/H3N2, A/H1N1 и B в эту смертность, и изменения в уровнях смертности после увеличения уровня вакцинации начиная с сезона 2016/17.

**Материалы и Методы**

*Данные*
Данные Росстата о месячной смертности от болезней органов дыхания и от болезней системы кровообращения были извлечены из [19]. Средние ежедневные уровни смертности на 100,000 человек по месяцам были получены из соответствующего месячного количества смертей, и данных о количестве населения [20] (население на данный месяц было оценено путем линейной интерполяции оценок на 1-е Января [20]). Известно, что грипп вносит вклад и в другие категории смертей (например от новообразований, а также от болезней, связанных с метаболизмом [3,4]); однако нужны данные с более высоким разрешением (например, недельные данные о смертности) для того, чтобы получить статистически значимый сигнал о связи между гриппом и этими категориями смертей.

Недельные данные о заболеваемости гриппом/ОРВИ на 10,000 человек в России доступны в [17]. Недельные данные о проценте респираторных образцов у пациентов с симптомами гриппа/ОРВИ, которые были ПЦР-положительны на каждый из основных подтипов гриппа (A/H3N2, A/H1N1, B) содержаться в [18].

### *Индексы циркуляции гриппа*

Только определенный процент пациентов с симптомами гриппа/ОРВИ заражен гриппом. Мы оценили недельные индексы циркуляции для каждого из основных подтипов гриппа (A/H3N2, A/H1N1, B) следующим образом

Недельный индекс циркуляции для данного подтипа гриппа =
(Уровень заболеваемости гриппом/ОРВИ) ∗ (% респираторных образцов, ПЦР − положительных на данный подтип гриппа)                              (1)

Как отмечалось в [3], недельные индексы циркуляции для каждого подтипа гриппа пропорциональны недельных уровням заболеваемости от данного подтипа гриппа в населении - по сути, эти индексы раны уровню заболеваемости от данного подтипа гриппа на 10,000 человек умноженными на чувствительность ПЦР-теста. Месячные индексы циркуляции для каждого из основных подтипов гриппа определяются как *взвешенная сумма* недельных индексов циркуляции данного подтипа гриппа, а именно, для данного месяца и подтипа гриппа, мы суммируем недельные индексы циркуляции гриппа данного подтипа для тех недель, которые пересекаются с данным месяцем, помноженные на количество общих дней у данных месяца и недели (7 дней если неделя полностью содержится в данном месяце), и в конце делим полученный результат на количество дней в данном месяце. Наконец, для того, чтобы связать индексы циркуляции гриппа со смертностью, ассоциированной с гриппом, мы учитываем время между заболеваемостью гриппом и смертью (в среднем, приблизительно неделя, [3,4]), и сдвигаем недельные индексы циркуляции на неделю вперед, а затем из этих сдвинутых недельных индексов подучаем месячные индексы, как описывается выше в этом абзаце.

Отношение между индексом циркуляции для данного подтипа гриппа и ассоциированной с этим подтипом гриппа смертностью может менятся во времени. Для гриппа B это связано прежде всего с циркуляцией гриппа B/Виктория и B/Ямагата. Распределение по возрасту у больных гриппом B/Виктория значительно моложе, чем у больных гриппом B/Ямагата [21,22]. Так как индексы циркуляции отражают циркуляцию гриппа в общем населении, а смертность от гриппа B в основном отражает смерти среди старших слоев населения, то соотношение между индексом циркуляции и смертностью может сильно отличаться для гриппа B/Виктория и B/Ямагата. Грипп B/Ямагата преобладал в циркуляции гриппа B в сезоны 2013/14, 2014/15, 2017/18, и 2018/19 (сезон определяется как период с сентября по июнь), а грипп B/Виктория преобладал в сезоны 2015/16 и 2016/17 [18]. Соответственно, мы разбиваем индекс циркуляции гриппа B на два: индекс в сезоны 2013/14, 2014/15, 2017/18, и 2018/19 (который мы назовем индекс B/Ямагата), и индекс в сезоны 2015/16 и 2016/17 (который мы назовем индекс B/Виктория). Для гриппа A/H3N2, сезоны 2014/15 и 2015/16 были охарактеризованы циркуляцией дрейф-вариантов, отличных от штамма гриппа, который присутствовал в вакцине; также в сезон 2014/15 значительная смертность от гриппа была зафиксирована в ряде стран, например [7]. В остальные сезоны (2013/14, 2016/17, 2017/18, 2018/19), большинство циркулирующим штаммов гриппа A/H3N2 были подобны вакцинному штаму. Соответственно, мы разбиваем индекс циркуляции гриппа A/H3N2 на два: индекс в сезоны 2014/15 и 2015/16 (который мы назовем индекс $A/H3N2^{\text{дрейф}}$), и индекс в сезоны 2013/14, 2016/17, 2017/18, 2018/19 (который мы назовем индекс $A/H3N2^{\text{вакцинный}}$). Наконец, мы отметим, что в течении всех 6-и изучаемых сезонов (2013/14 до 20189/19), большинство циркулирующих штаммов гриппа A/H1N1 совпадали с вакцинным штаммом. На Рис. 2 изображены месячные индексы циркуляции гриппа A/H3N2, A/H1N1 и B в период с 07/2013 до 07/2019, включая разбиения на B/Ямагата и B/Виктория, и разбиения на $A/H3N2^{\text{дрейф}}$ и $A/H3N2^{\text{вакцинный}}$. Мы также отметим, что аналогичные индексы циркуляции гриппа могут служить индикаторами уровня циркуляции гриппа в реальное время в будующие сезоны гриппа.

*Статистический анализ*

Мы связываем месячные индексы циркуляции гриппа с месячными уровнями смертности в России (отдельно для смертности от болезней органов дыхания и болезней системы кровообращения) следующим образом: Пусть С(м) будет уровень соответствующей смертности в месяц м (м = 1 для 07/2013; м=73 для 07/2019), и $A/H3N2^{\text{дрейф}}$(м), $A/H3N2^{\text{вакцинный}}$(м), $A/H1N1$(м), B/Ямагата(м), B/Виктория(м) будут соответствующие индексы циркуляции гриппа в месяц м. Тогда

С(м) = $\beta_0$ + $\beta_1 \cdot A/H3N2^{\text{дрейф}}$(м) + $\beta_2 \cdot A/H3N2^{\text{вакцинный}}$(м)(м) + $\beta_3 \cdot A/H1N1$(м) + $\beta_4 \cdot$ B/Ямагата(м) + $\beta_5 \cdot$ B/Виктория(м) + Базовый уровень(м) + Тренд(м) + Белый шум

(2)

"Базовый уровень(м)" отражает ожидаемый уровень смертности от данного класса болезней, не связанной с гриппом в месяц м – предполагается, что этот уровень периодичен с годовой периодичностью. Мы моделируем "Базовый уровень(м)" как

Базовый уровень(м)= $\beta_6 \cdot \cos\left(\frac{2\pi \text{м}}{12}\right) + \beta_7 \cdot \sin\left(\frac{2\pi \text{м}}{12}\right) + \beta_8 \cdot$ Янв

Здесь переменное "Янв" равно 1 для месяца Январь, и нулю для других месяцев. Это переменное включено в регрессионную модель потому, что месячные данные о смертности [19] являются оперативными, и данные, не внесенные в систему в течении календарного года переносятся на Январь следующего года [23]. Более того, наличие Январского эффекта на смертность от болезней системы кровообращения в России было выявлено в [24]. Мы также отметим, что предыдущие работы показали, что выбор модели для базовых уровней смертности имеет ограниченное влияние на оценку уровня смертности, ассоциированной с гриппом (дополнительная информация в статье [3]).

Тренд(м) отражает временные тенденции в смертности от данного класса болезней. Тренд(м) моделируется как многочлен низкой степени от месяца м.

Мы используем информационный критерий Акаике (AIC) для выбора параметров в модели в уравнении 2. При каждом шаге, параметр, чъе удаление из модели приводит к максимальному уменьшнию оценки AIC опускается; это повторяется до тех пор, пока удаление любого из оставшихся параметров не приводит к уменьшению оценки AIC. Это приводит к соответствующим моделям для смертности от болезней органов дыхания, и болезней системы кровообращения

*Болезни органов дыхания*

$$С_{дыхание}(м) = \beta_0 + \beta_1 \cdot A/H3N2^{дрейф}(м) + \beta_2 \cdot A/H3N2^{вакцинный}(м) + \beta_3 \cdot A/H1N1(м) + \beta_4 \cdot В/Ямагата(м) + \beta_5 \cdot \sin\left(\frac{2\pi м}{12}\right) + \beta_6 \cdot Янв(м) + \beta_7 \cdot м + \beta_8 \cdot м^2 + \beta_9 \cdot м^3 \qquad (3)$$

*Болезни системы кровообращения*

$$С_{кровообращение}(м) = \beta_0 + \beta_1 \cdot A/H3N2^{дрейф}(м) + \beta_2 \cdot A/H3N2^{вакцинный}(м) + \beta_3 \cdot A/H1N1(м) + \beta_4 \cdot В/Ямагата(м) + \beta_5 \cdot \sin\left(\frac{2\pi м}{12}\right) + \beta_6 \cdot Янв(м) + \beta_7 \cdot м + \beta_8 \cdot м^2 \qquad (4)$$

Грипп В/Виктория был исключен из модели посредством информационного критерия Акаике как для смертности от болезней органов дыхания, так и для смертности от болезней системы кровообращения (см. Результаты). Остальные индексы циркуляции гриппа вошли в обе модели.

**Результаты**

Рис. 1 отображает ежедневный уровень смертности от болезней органов дыхания на миллион человек, и от болезней системы кровообращения на 100,000 человек в РФ по месяцам в период с 07/2013 до 07/2019. Рис. 2 отображает месячные индексы циркуляции гриппа $A/H3N2^{дрейф}(м), A/H3N2^{вакцинный}(м), A/H1N1(м), В/Ямагата(м),$

В/Виктория(м) в период с 07/2013 до 07/2019. Отметим, что есть хорошее визуальное соответствие между периодами высокой циркуляции гриппа и скачками в смертности, за исключением циркуляции гриппа В/Виктория в сезон 2016/17. Отсутствие соответствующего сигнала в кривой смертности при высокой циркуляции гриппа В/Виктория может быть объяснено тем, что этот грипп относительно редко поражает немолодых людей [21,22].

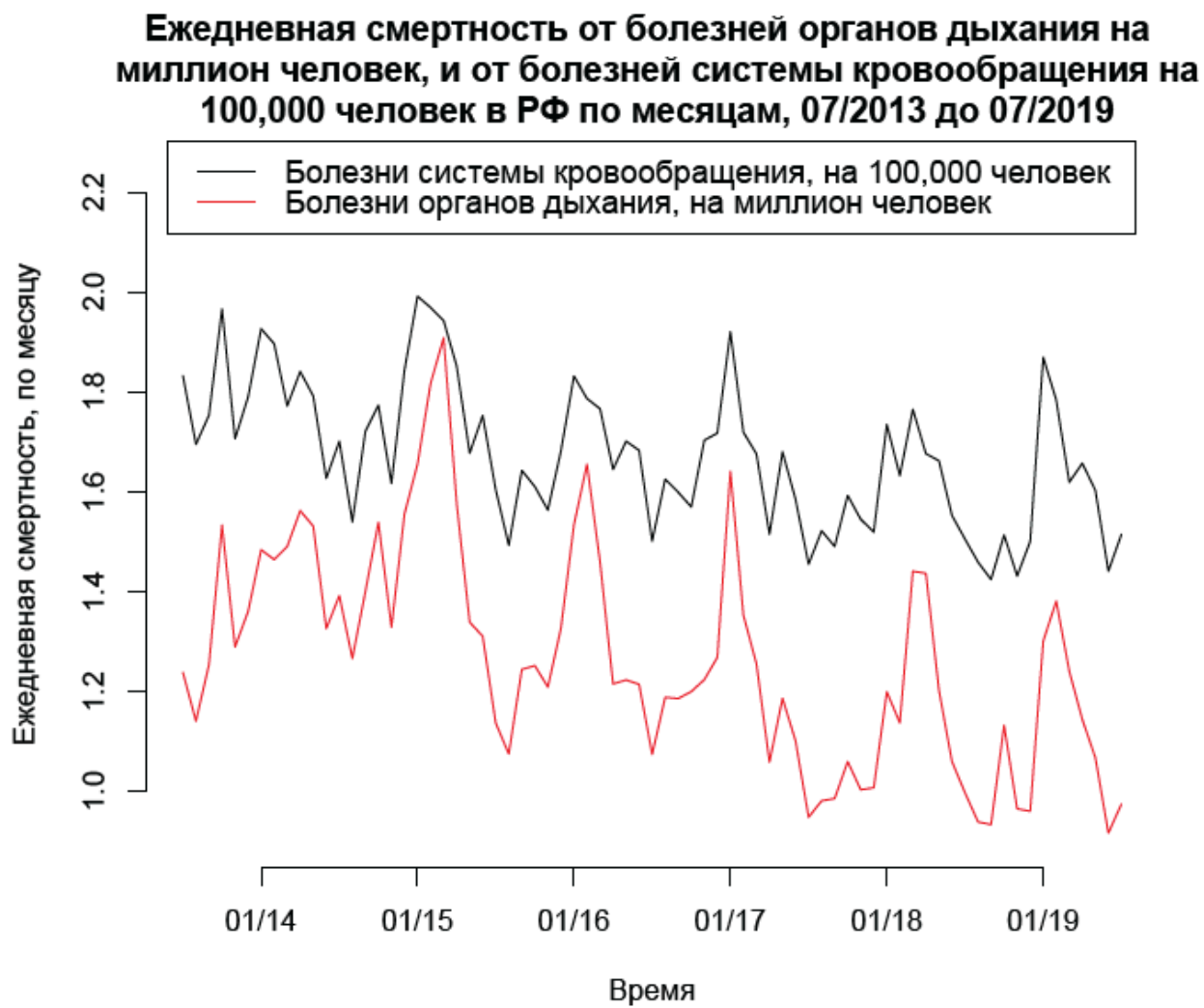

**Рис. 1** Средний ежедневный уровень смертности от болезней органов дыхания на миллион человек и от болезней системы кровообращения на 100,000 человек в РФ по месяцам, 07/2013 до 07/2019.

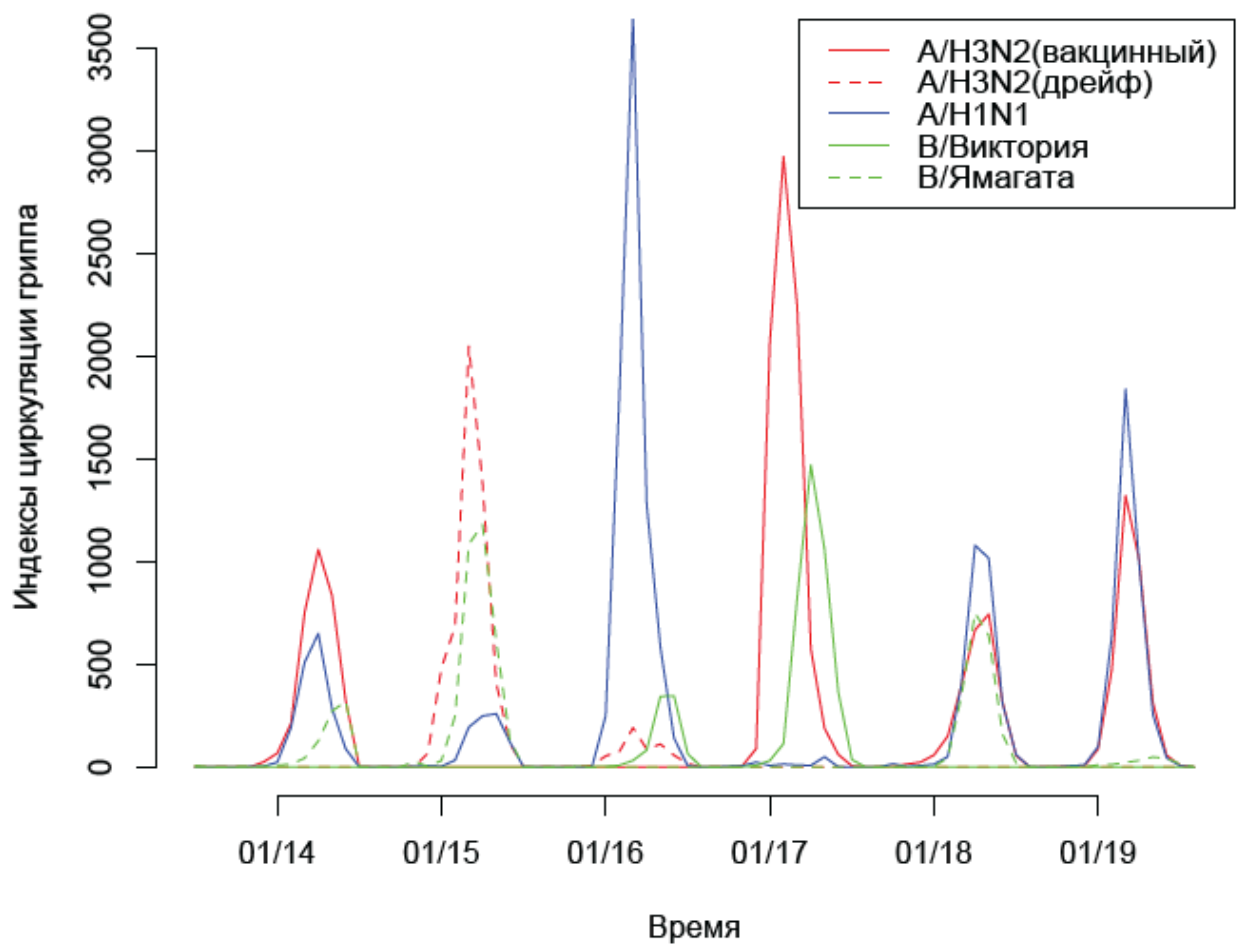

**Рис. 2**: Месячные индексы циркуляции гриппа A/H3N2, A/H1N1 и B (уравнение 1) в РФ в период с 07/2010 до 07/2019.

Рис. 3 отображает результаты модели для смертности от болезней органов дыхания, заданной уравнением 3. Рис. 4 отображает результаты модели для смертности от болезней системы кровообращения, заданной уравнением 4. Результаты моделей достаточно последовательно отражают уровни смертности, особенно для болезней органов дыхания. Это дает подтверждение уместность структуры нашей модели, т.е. месячные уровни смертности большей частью описываются вкладом гриппа в дополнении к регулярному образцу (базовые уровни + тренд).

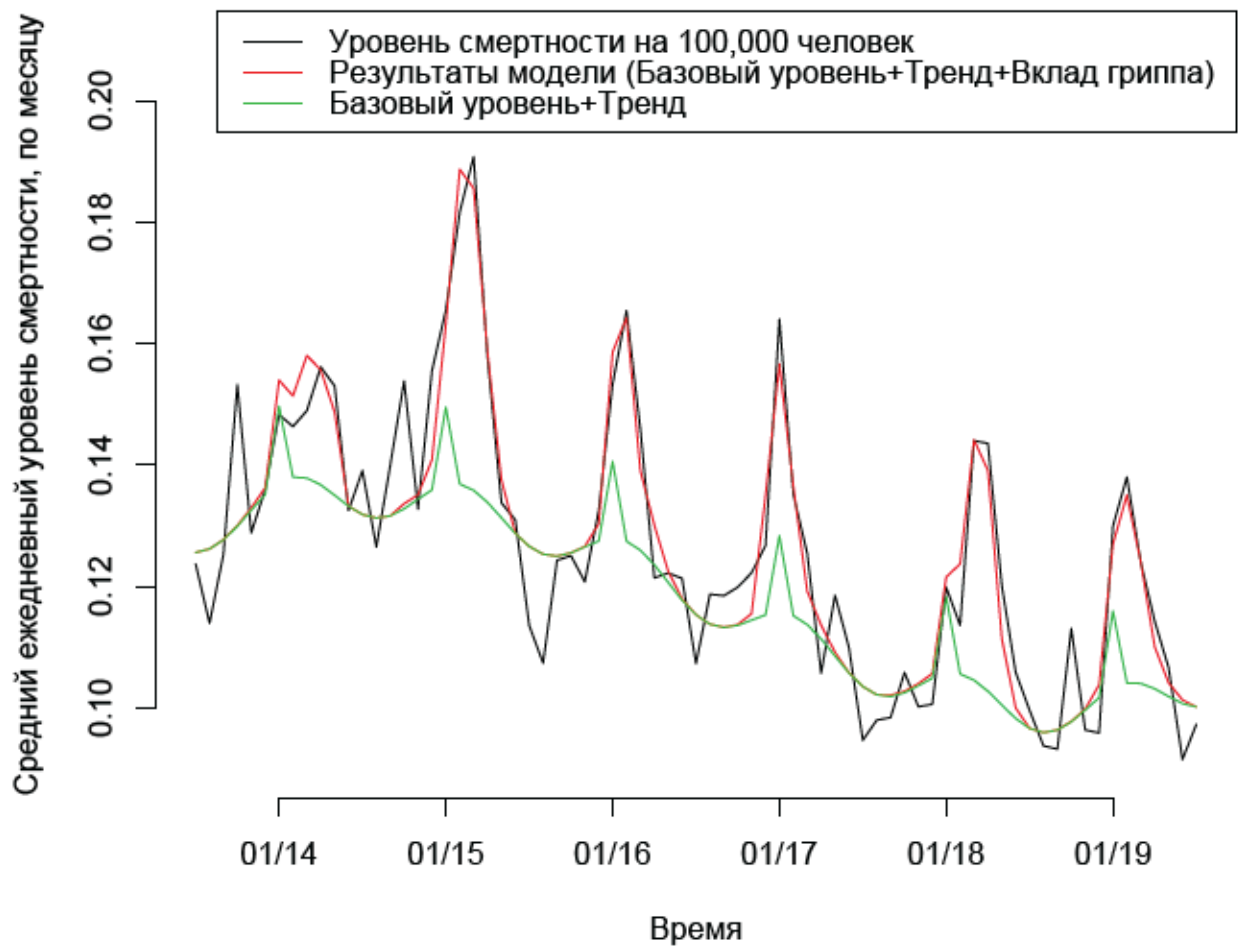

**Рис. 3** Средний ежедневный уровень смертности от болезней органов дыхания 100,000 человек в РФ по месяцам, 07/2013 до 07/2019 (черная кривая), результаты модели (красная кривая), и базовый уровень смертности, не связанной с гриппом + тренд (зеленая кривая). Вклад гриппа в смертность равен разнице между красной и зеленой кривыми.

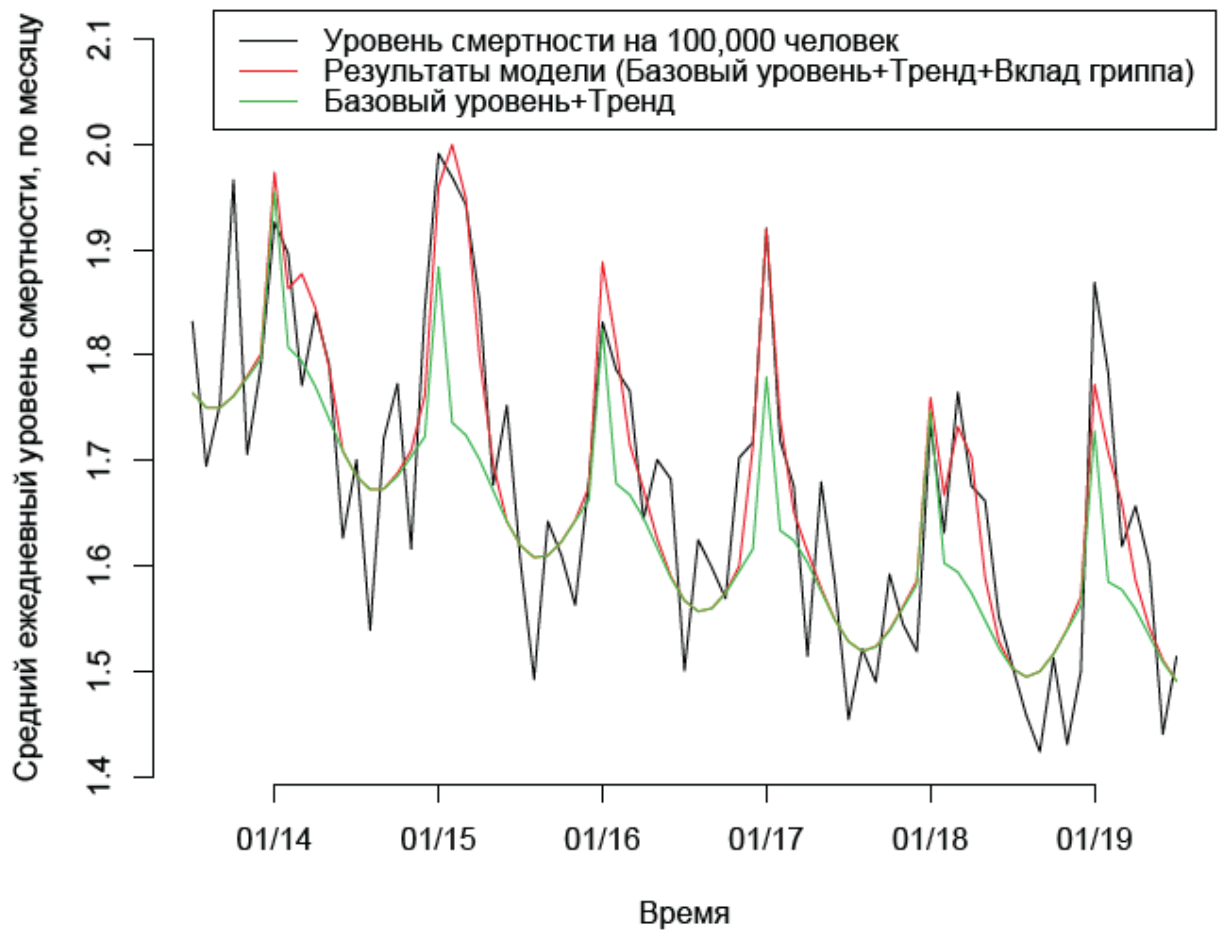

**Рис. 4** Средний ежедневный уровень смертности от болезней системы кровообращения 100,000 человек в РФ по месяцам, 07/2013 до 07/2019 (черная кривая), результаты модели (красная кривая), и базовый уровень смертности, не связанной с гриппом + тренд (зеленая кривая). Вклад гриппа в смертность равен разнице между красной и зеленой кривыми.

В таблице 1 представлены оценки годового вклада гриппа в смертность от болезней органов дыхания, и болезней системы кровообращения для каждого из сезонов гриппа 2013/14 по 2018/19 (сезон определяется как период с сентября по июнь), а также среднегодовые значения для соответствующего вклада гриппа в смертность за этот период времени (6 сезонов). Мы оценили, что в среднем 17636 (95% ДИ (9482,25790)) годовых смертей от болезней системы кровообращения и 4179

(3250,5109) смертей от болезней органов дыхания в сезоны с 2013/14 до 2018/19 были ассоциированы с гриппом. Наибольшая смертность как от болезней системы кровообращения (32298 (18071,46525)), так и от болезней органов дыхания (6689 (5019,8359)), ассоциированная с гриппом была оценена в сезон 2014/15, когда дрейфовые варианты гриппа A/H3N2 и B/Ямагата циркулировали в России. В сезон 2014/15 высокая смертность, связанная с циркуляцией гриппа A/H3N2 и B/Ямагата быда зафиксирована и в ряде других стран [7]. Уровень вакцинации против гриппа в России резко возрос начиная с сезона 2016/17. По сравнению с сезонами 2013/14 до 2015/16, в сезоны 2016/17 до 2018/19 смертность, ассоциированная с гриппом упала на 16.1%, или 3809 ежегодных смертей от болезней системы кровообращения и болезней органов дыхания. Наконец, мы отметим высокий уровень смертности от болезней системы кровообращения, ассоциированной с гриппом. В частности, уровень смертности от болезней системы кровообращения, ассоциированной с гриппом превышает уровень смертности от болезней органов дыхания, ассоциированной с гриппом в 4.22 раза. В США отношение между уровнем смертности от болезней системы кровообращения и уровнем смертности от болезней органов дыхания, ассоциированной с гриппом оценивается в 1.35 [7]. При этом, сооотношение общего количества смертей от болезней системы кровообращения к количеству смертей от болезней органов дыхания в России (14-к-1 в 2018-м г., [20]) значительно превышает сооотношение общего количества смертей от болезней системы кровообращения к количеству смертей от болезней органов дыхания в США (3.08-к-1 в 2017-м г., [25]).

| Сезон | Болезни системы кровообращения | Болезни органов дыхания |
| --- | --- | --- |
| 2013/14 | 12364 (5497,19230) | 3116 (2329,3902) |
| 2014/15 | 32298 (18071,46525) | 6689 (5019,8359) |
| 2015/16 | 13128 (1046,25211) | 3565 (2171,4959) |
| 2016/17 | 17462 (4376,30548) | 3540 (2045,5035) |
| 2017/18 | 17463 (6294,28632) | 4887 (3606,6167) |
| 2018/19 | 13101 (4760,21442) | 3280 (2330,4229) |
| Годовое среднее | 17636 (9482,25790) | 4179 (3250,5109) |

**Таблица 1**: Количество смертей от болезней системы кровообращения и болезней органов дыхания, ассоциированных с гриппом в сезоны с 2013/14 до 2018/19 в РФ, и годовое среднее для соответствующих 6-и сезонов гриппа.

В таблице 2 представлены оценки среднего годового количества смертей от болезней системы кровообращения и болезней органов дыхания, ассоциированных с основными подтипами гриппа в сезоны с 2013/14 до 2018/19 в РФ. Мы оценили, что в тот период, среди смертей, ассоциированных с гриппом, 51.8% смертей от болезней системы кровообращения и 37.2% смертей от болезней органов дыхания были связаны с гриппом A/H3N2 (Таблица 2 и 1); 23.4% смертей от болезней системы кровообращения и 29.5% смертей от болезней органов дыхания были связаны с гриппом A/H1N1; 24.9% смертей от болезней системы кровообращения и 33.3% смертей от болезней органов дыхания были связаны с гриппом B.

| Подтип гриппа | Болезни системы кровообращения | Болезни органов дыхания |
| --- | --- | --- |
| A/H3N2 | 9140 (2608,15672) | 1553 (805,2301) |
| A/H1N1 | 4100 (-156,8355) | 1233 (746,1721) |
| B | 4396 (-760,9553) | 1393 (799,1987) |

**Таблица 2**: Среднее годовое количество смертей от болезней системы кровообращения и болезней органов дыхания, ассоциированных с гриппом A/H3N2, A/H1N1 и B в сезоны с 2013/14 до 2018/19 в РФ.

**Обсуждение**

Известно, что циркуляция гриппа связана с существенным бременем смертности в Северном полушарии, например [3-9]. В то же время, информация о смертности, ассоциированной с гриппом в РФ ограничена, и в основном базируется на данных о смертях с диагностированным гриппом [2,12-15]. Эти смерти представляют только малую долю всех смертей, ассоциированных с гриппом, и не позволяют оценить вклад разных подтипов гриппа в смертность, связанную с гриппом, а также уровень смертности от разных болезней, ассоциированной с гриппом (Введение). В этой статье мы применили ранее разработанную методологию, которая уже была использована для оценки уровня смертности, ассоциированной с гриппом в ряде стран [3,4,6-9] для того, чтобы оценить уровень смертности от болезней органов дыхания и болезней системы кровообращения в РФ в сезоны гриппа с 2013/14 до 2018/19. Мы оценили в среднем 17636 (95% ДИ (9482,25790)) годовых смертей от болезней системы кровообращения и 4179 (3250,5109) смертей от болезней органов дыхания в сезоны с 2013/14 до 2018/19, ассоциированных с гриппом. Отметим, что высокий уровень смертности от болезней системы кровообращения, ассоциированной с гриппом совместим с общим высоким уровнем смертности от болезней системы кровообращения в РФ, а также связью между инфекцией гриппа и рядом сердечно-сосудистых осложнений [10,11]. Самый большой вклад как в смертность от болезней системы кровообращения, так и в смертность от болезней органов дыхания внес грипп A/H3N2, что соответствует оценкам и в других странах [3,5,7,9]. Грипп A/H1N1 и B (в основном B/Ямагата, что согласуется с рядом исследований [26,21,22]) тоже внесли существенный вклад в смертность. К тому же, значительное увеличение уровня вакцинации против гриппа в РФ начиная с сезона 2016/17 было связано с уменьшением уровня смертности от гриппа (в среднем на 16.1%, или почти 4,000 смертей в год). Мы надеемся, что эта работа поможет планировать усилия по уменьшению тяжелых последствий, связанных с эпидемиями гриппа. В частности, наши результаты являются свидетельством в поддержку дополнительного увеличения уровня вакцинации против гриппа, особенно среди людей с сердечно-сосудистыми заболеваниями и пожилых людей. Вклад гриппа B/Ямагата в смертность, особенно в сезоны, когда этот вид гриппа не содержался в трехвалентной вакцине, является свидетельством в поддержку использования четырехвалентых вакцин против гриппа. Отметим также, что в предыдущий сезон 2018/19, циркуляция гриппа B в России была минимальной, что увеличивает шансы значительной эпидемии гриппа B в сезон 2019/20; также, грипп B/Ямагата не содержится в трехвалентной вакцине, рекомендованной в сезон 2019/20 [27]. Мы также отметим, что наши результаты являются свидетельством в поддержку применения антивирусных препаратов, особенно для представителей групп риска, таких как люди с сердечно-сосудистыми и другими хроническими заболеваниями и пожилые люди, включая применение в амбулаторных условиях при симптомах гриппа в периоды высокого уровня циркуляции гриппа. Наконей мы отметим, что индексы циркуляции гриппа, определенные в этой статье могут служить индикаторами уровня циркуляции гриппа в реальное время в будующие сезоны гриппа.

Наша статья имеет несколько ограничений. Основное ограничение заключается в том, что мы использовали месячные, оперативные данные о смертности [23], а не недельные данные. Данные о смертности, стратифицированные по неделям/возрастным группам помогут лучше понять эпидемиологию смертности, связанной с циркуляцией гриппа в России. В частности, из-за относительно небольшого количества месячных данных, мы использовали тригонометрическую модель для уровней базовой смертности, не связанной с гриппом. Предыдущие работы показали, что выбор модели для базовых уровней смертности имеет ограниченное влияние на оценку уровня смертности, ассоциированной с гриппом (дополнительная информация в статье [3]).

**Выводы**

Несмотря на определенные ограничения, наша работа представляет оценку уровня смертности от болезней системы кровообращения и болезней органов дыхания, ассоциированной с гриппом в Российской Федерации во время последних 6-и сезонов гриппа (с 2013/14 до 2018/19). Наши оценки подтверждают и квантифицируют наблюдаемую связь между высокой циркуляцией гриппа и повышением уровня смертности от болезней системы кровообращения и болезней органов дыхания. Мы нашли значительный вклад гриппа, особенно гриппа A/H3N2, но также гриппа A/H1N1 и B в смертность, особенно от болезней системы кровообращения. Мы также нашли эффект недавнего увеличения уровня вакцинации на смертность от гриппа (уменьшение в среднем почти на 4000 смертей в год начиная с сезона 2016/17). Наши результаты являются свидетельством в поддержку дополнительного увеличения уровня вакцинации против гриппа, особенно среди людей с сердечно-сосудистыми заболеваниями и пожилых людей, использования четырехвалентых вакцин против гриппа, и применения антивирусных препаратов, особенно для представителей групп риска, таких как люди с сердечно-сосудистыми и другими хроническими заболеваниями и пожилые люди, в периоды высокого уровня циркуляции гриппа. Мы также надеемся, что эта работа послужит стимулом для более подробного изучения эпидемиологии тяжелых последствий эпидемий гриппа в России, основанного на более гранулярных данных, в целях планирования усилий по уменьшению смертности и других тяжелых последствий, связанных с эпидемиями гриппа.

**Ссылки**